\documentclass[aps,prb,twocolumn,superscriptaddress,showpacs]{revtex4}
\usepackage[T1]{fontenc}
\usepackage[latin9]{inputenc}
\usepackage{amsmath}
\usepackage{amssymb}
\usepackage{graphicx}
\usepackage{esint}
\usepackage{bm}
\usepackage{amsfonts,color}

\renewcommand{\vec}[1]{\bm{#1}}

\DeclareMathAlphabet{\mathsfsl}{OT1}{cmss}{m}{sl}

\begin{document}

\title{Numerical study of the giant nonlocal resistance in spin-orbital coupled graphene}

\author{Zibo Wang}
\affiliation{International Center for Quantum Materials, School of Physics, Peking University, Beijing 100871, China}
\affiliation{Microsystems and Terahertz Research Center, China Academy of Engineering Physics, Chengdu, Sichuan 610200, China}

\author{Haiwen Liu}
\affiliation{Center for Advanced Quantum Studies, Department of Physics, Beijing Normal University, Beijing 100875, China }

\author{Hua Jiang}
\thanks{\texttt{jianghuaphy@suda.edu.cn}}
\affiliation{Department of Physics, Soochow University, Suzhou 215006, China}

\author{X. C. Xie}
\affiliation{International Center for Quantum Materials, School of Physics, Peking University, Beijing 100871, China}
\affiliation{Collaborative Innovation Center of Quantum Matter, Beijing 100871, China}

\date{\today}

\begin{abstract}
Recent experiments find the signal of giant nonlocal resistance $R_{NL}$ in H-shaped graphene samples due to the spin/valley Hall effect. Interestingly, when the Fermi energy deviates from the Dirac point, $R_{NL}$ decreases to zero much more rapidly compared with the local resistance $R_L$, and the well-known relation of $R_{NL}\propto R_L^3$ is not satisfied.
In this work, based on the non-equilibrium Green's function method,
we explain such transport phenomena in the H-shaped graphene with Rashba spin-orbit coupling.
When the Fermi energy is near the Dirac point, the nonlocal resistance is considerably large and is much sharper than the local one.
Moreover, the relationship between the Rashba effect and the fast decay of $R_{NL}$ compared with $R_L$ is further investigated.
We find that the Rashba effect does not contribute not only to the fast decay but also to the peak of $R_{NL}$ itself.
Actually, it is the extremely small density of states near the Dirac point that leads to the large peak of $R_{NL}$,
while the fast decay results from the quasi-ballistic mechanism.
Finally, we revise the classic formula $R_{NL}\propto R_L^3$
by replacing $R_{NL}$ with $R_{Hall}$, which represents the nonlocal resistance merely caused by the spin Hall effect,
and the relation holds well.
\end{abstract}

\date{\today}

\maketitle

\section{Introduction}

Nonlocal measurement refers to the detection of a voltage signal outside the path, along which charge current is expected to flow.
One way to generate the nonlocal voltage is to modify the charge current path away from the classic Ohmic mode\cite{Andreas,Chang, Parameswaran}.
For instance, in the quantum Hall regime, the current transports along edges while the bulk is insulating\cite{McEuen}.
Another important way for the generation of the nonlocal voltage is to induce current with other degrees of freedom (i.e. spin/valley),
so that the current direction will deviate from the exciting field.
Since the nonlocal voltage always originates from nontrivial physics that is not easy to detect directly,
the nonlocal measurement has now become a powerful tool to discover such kinds of electromagnetic phenomena in many novel materials
\cite{Abanin1,Balakrishnan,Gorbachev,Shimazaki,Sui,Michihisa}.

Spin Hall effect (SHE) is a phenomenon arising from the spin-orbit coupling
in which charge current passing through a sample leads to spin transport in the transverse direction
\cite{Hirsch,Murakami,Sinova,Kato,Kimura,Brune}.
Since only the electron spin, rather than the electron charge, accumulates during the spin transport process, it is always difficult to observe the SHE with local measurements.
Fortunately, with the method of nonlocal measuring,
a large nonlocal resistance $R_{NL}$ was reported near the Dirac point in H-shaped graphene sample
\cite{Abanin1,Balakrishnan},
which confirms the existence of the SHE.
Moreover, the nonlocal measurement was also used to detect valley Hall effect (VHE)
\cite{Gorbachev,Shimazaki,Sui,Michihisa}, whose transport mechanism is similar to the SHE,
and a giant nonlocal resistance can be observed as well.

Abanin et al. developed a theory to discuss the origin of the above-mentioned nonlocal resistance\cite{Abanin2}.
In their paper, they demonstrated that the relationship between the nonlocal resistance $R_{NL}$ and the local resistance $R_L$ can be described by a simple function of
$R_{NL}\propto \sigma_{xy}^2R_L^3$, where $\sigma_{xy}$ is the spin Hall conductance.
However, this equation remains difficult to explain some experimental results.
Specifically, besides the giant peak of $R_{NL}$, people also find another interesting phenomenon that,
compared with $R_L$, $R_{NL}$ decays much more rapidly when Fermi energy deviates from the Dirac point.
We consider Fig.1(b) of Ref.[\onlinecite{Gorbachev}] as an example. When $V_g=1$V,
the red line of $R_{NL}$ has already collapsed to zero,
whereas the black line of $R_L$ is still finite.
This novel phenomenon seems to be inconsistent with Ref.[\onlinecite{Abanin2}],
because the zero value of $R_{NL}$ can not be proportional to $R_L^3$,
which deserves further explanations.

Recently, with an extrinsic perpendicular electric field,
Chen's group at Peking University also detected a nonlocal voltage signal in an H-shaped graphene sample\cite{Chen}.
When the Fermi energy deviates from the Dirac point,
the nonlocal resistance shows analogous behavior that $R_{NL}$ decreases to zero much more quickly than $R_L$.
This experimental work motivates us to study the nonlocal resistance numerically
by considering standard monolayer graphene with an extrinsic Rashba effect.
Firstly, in an H-shaped four-terminal system,
we obtain $R_L$ and $R_{NL}$ by means of the non-equilibrium Green's function method,
where the numerical results exhibits the same properties as the experimental findings, namely,
a giant peak and an obviously fast decay of the nonlocal resistance by tuning the Fermi energy.
Secondly, we find that $R_{NL}$ can be negative in a certain region.
This phenomenon implies the existence of the quasi-ballistic transport mechanism,
which was observed previously\cite{Mihajlovic}.
Therefore, we conclude that $R_{NL}$ consists of three parts: $R_{ballistic}$ due to the ballistic mechanism,
$R_{classic}$ from the classic diffusion and $R_{Hall}$ from the SHE.
Since the negative value of $R_{ballistic}$ locates around the Dirac point,
it is possible that the quasi-ballistic mechanism contributes to the fast decay of $R_{NL}$.
Thirdly, in order to further investigate the relationship between this fast decay and the Rashba effect,
we study a six-terminal system and obtain the nonlocal resistance $R_{Hall}$ which is only caused by the SHE.
Surprisingly, the results show that $R_{Hall}$ equals to zero at the Dirac point.
Since the SHE always makes a nonnegative contribution to $R_{NL}$,
we conclude that there is not any relationship between the Rashba effect and the fast decay of $R_{NL}$.
In fact, it is the extremely small density of states (DOS) near the Dirac point and the quasi-ballistic mechanism
that lead to the large peak and the fast decay of $R_{NL}$, respectively.
Moreover, the Rashba effect itself actually plays a negative role in the fast decay of $R_{NL}$.
Finally, with the spin Hall conductance $\sigma_{xy}$ calculated in a four-terminal system,
we modify the previous theoretical formula $R_{NL}\propto \sigma_{xy}^2R_L^3$ by replacing $R_{NL}$ with $R_{Hall}$,
and find that this revised formula holds well for the present case.

The rest of this paper is organized as follows.
In Sec.II, we numerically calculate the local and nonlocal resistance in an H-shaped four-terminal system.
Then, in Sec.III, we study a six-terminal system to obtain the nonlocal resistance which is merely caused by the SHE,
and compare it with the previous theoretical prediction.
Finally, a conclusion is presented in Sec.IV.

\section{H-shaped four-terminal system to study nonlocal resistance}

\begin{figure}[h]
\includegraphics [width=\columnwidth]{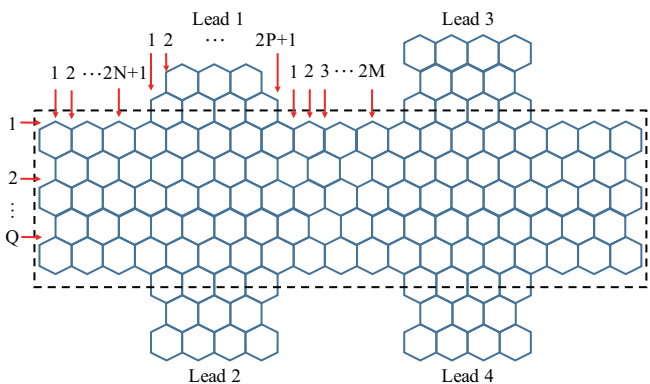}
\caption{(Color online)
The schematic diagram of the proposed H-shaped four-terminal system.
The current is injected into lead 1 and flows out of lead 2.
The voltage signal is obtained between leads 1 and 2 for the local resistance,
and is derived between leads 3 and 4 for the nonlocal resistance.
The rectangle, denoted by the black dashed line, is the center region.
}\label{fig:fourdiagram1}
\end{figure}

We first consider an H-shaped four-terminal system to simulate the nonlocal measurement\cite{foot1},
where the schematic diagram is shown in Fig.~\ref{fig:fourdiagram1}.
Due to the strong external electric field, the tight-binding Hamiltonian can be written as:
\begin{eqnarray}
H=\sum_i\epsilon_ic_i^\dagger c_i+t\sum_{\langle ij\rangle} c_i^\dagger c_j+i\lambda_R\sum_{\langle ij\rangle}c_i^\dagger(\vec{s}\times\hat{\vec{d}}_{ij})_zc_j,
\label{eq:Hamiltonian}
\end{eqnarray}
where $c_i^\dagger$ and $c_i$ are the creation and annihilation operators, respectively, at site $i$,
$\epsilon_i$ is the on-site energy,
and $\lambda_R$ is the strength of the external Rashba effect.
The on-site energies of the four terminals are chosen as $\epsilon_{1,2,3,4}=\epsilon_{UD}$ to simulate metallic leads,
which can be controlled by the gate voltage.
The disorder only exists in the central region
and is modeled by Anderson disorder with the on-site energies being uniformly distributed in $[-w/2,w/2]$,
where $w$ is the disorder strength.
The schematic diagram of the central region is described by parameters $M,N,P$ and $Q$.
For instance, figure~\ref{fig:fourdiagram1} shows a system with $M=3,N=2,P=4$ and $Q=3$.
Therefore, the system has $2Q[(2N+2)+(2P+1)+(2M+1)+(2P+1)+(2N+2)]$ carbon atoms in total.

The current flowing through the four-terminal system can be calculated from
the Landauer-B\"{u}ttiker formula: $I_i=\frac{e}{h}\sum_j\int {\rm d}\epsilon T_{ij}(\epsilon)[f_i(\epsilon)-f_j(\epsilon)]$,
where $f_i(\epsilon)=1/\{ 1+{\rm exp}[(\epsilon-\mu_i)/k_BT]\}$ is the Fermi distribution function in the $i$th lead.
After applying a small electric field between the leads, the chemical potential of
lead i becomes $\mu_i=E_F+eV_i$.
At zero temperature, the former formula can be simplified as:
\begin{eqnarray}
I_i & = & \frac{e}{h}\sum_j\int {\rm d}\epsilon T_{ij}(\epsilon)[\theta(E_F+eV_i-\epsilon)-\theta(E_F+eV_j-\epsilon)]\nonumber\\
    & = & \frac{e}{h}\sum_j\int^{E_F+eV_i}_{E_F+eV_j}{\rm d}\epsilon T_{ij}(\epsilon) \nonumber \\
    & = & \frac{e^2}{h}\sum_j T_{ij}(E_F)(V_i-V_j),
\label{eq:current}
\end{eqnarray}
where $T_{ij}(\epsilon)={\rm Tr}(\Gamma_iG^r\Gamma_jG^a)$ is the transmission coefficient
with the linewidth functions $\Gamma_i=i(\Sigma^r_i-\Sigma^a_i)$,
and the Green's function $G^r(\epsilon)=[G^a(\epsilon)]^\dagger=1/(\epsilon-H_{center}-\Sigma^r_i-\Sigma^r_j)$.
Here, $\Sigma^r_i$ is the retarded self-energy due to the coupling to the lead $i$,
and $H_{center}$ is the Hamiltonian in the central region.
After obtaining $T_{ij}$, the current $I_1$ and the voltage $V_{3,4}$ can be further deduced according to Eq.~(\ref{eq:current})
under the conditions of $V_1=-V_2=V$ and $I_3=I_4=0$.
At last, the local and nonlocal resistance can be calculated\cite{book,Jiang}.
To be specific, the local resistance is defined as: $R_L=(V_1-V_2)/I_1$,
and the nonlocal resistance is defined as: $R_{NL}=(V_3-V_4)/I_1$.

Throughout this work, we take the nearest hopping energy $t\approx2.75{\rm eV}$ as the energy unit.
In the following calculation, the value of the on-site energy $\epsilon_{UD}$
is testified in the energy bands of zigzag ribbons, in order to simulate four metallic leads.
The size parameters $M$, $N$, $P$ and $Q$ are chosen as $M=30$, $N=30$, $P=100$ and $Q=100$.
This indicates that the size of the model we calculated is about $73nm\times42nm$.\cite{footnote}
In the presence of Anderson disorder, the resistance has been averaged over 50 times.

\begin{figure}[htbp!]
\includegraphics [width=\columnwidth]{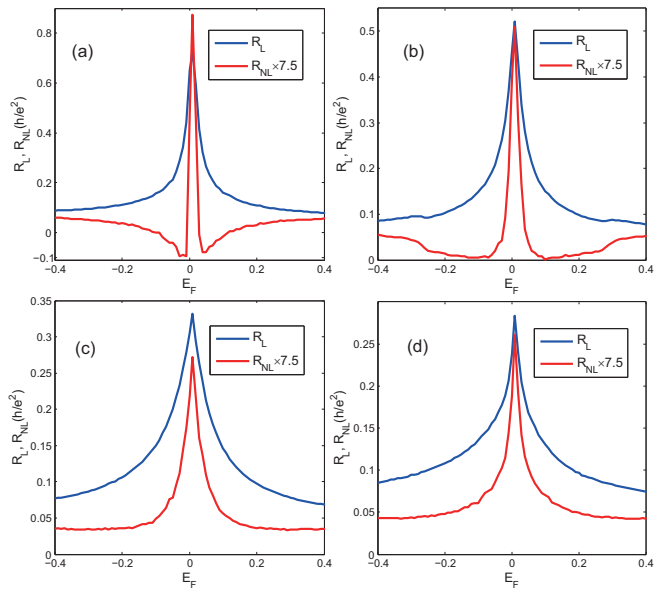}
\caption{(Color online)
The local $R_L$ and nonlocal resistance $R_{NL}$ are drawn in the blue and red line, respectively.
(a) $\lambda_R=0$; (b) $\lambda_R=0.1$; (c) $\lambda_R=0.2$; (d) $\lambda_R=0.3$.
The Anderson disorder strength is chosen as $w=1$.
In order to make the comparison clear enough, the value of $R_{NL}$ is amplified by $7.5$ times.
}\label{fig:nonlocal1}
\end{figure}

In Fig.~\ref{fig:nonlocal1}, we show the local resistance (blue lines) and the nonlocal resistance (red lines) as a function of the Fermi energy $E_F$.
From Fig.~\ref{fig:nonlocal1}(a) to \ref{fig:nonlocal1}(d), the Rashba spin-orbit strength $\lambda_R$ increases
from $\lambda_R=0$ to $\lambda_R=0.3$, while the other parameters remain unchanged.
As we can see, there are three main features in these four figures.

\begin{figure*}[htbp!]
\scalebox{2}{\includegraphics [width=\columnwidth]{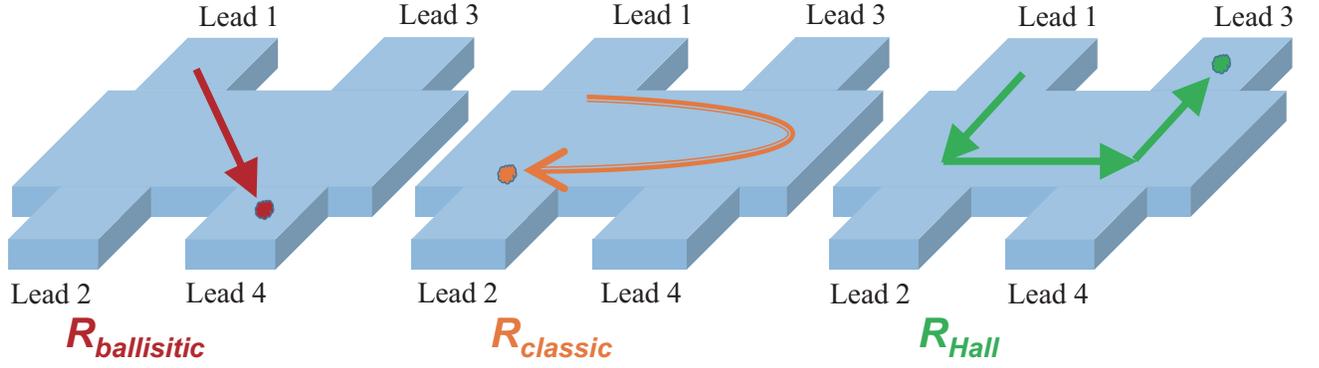}}
\caption{(Color online)
The schematic diagram for three kinds of transport mechanisms.
The electron current is injected into lead 1 and then flows out of lead 2.
The red line stands for the quasi-ballistic transport mechanism $R_{ballistic}$,
which makes a negative contribution to $R_{NL}$.
The yellow line represents the classic diffusion $R_{classic}$.
And the green line denotes the spin Hall transport $R_{Hall}$.
Arrows indicate the direction of the electron current which is injected into lead 1.}\label{fig:fourdiagram2}
\end{figure*}

Firstly, the nonlocal resistance exhibits negative value in Fig.~\ref{fig:nonlocal1}(a).
Such ``negative'' means that when the current flows from lead 1 to lead 2,
the voltage detected on lead 3 is surprisingly lower than that on lead 4.
A similar behavior can also be found in Fig.~\ref{fig:nonlocal1}(b)
that a pair of dips exist at about $E_F=\pm0.1$,
though their values are not negative.
This phenomenon can be explained by the quasi-ballistic transport mechanism
which was first predicted in experiments\cite{Mihajlovic}.
Specifically, the charge carriers injected into lead 1 can flow directly to lead 4 without returning back to lead 2.
This indicates that we can detect a positive voltage on lead 4 and a negative voltage on lead 3,
which means the value of the nonlocal resistance is negative.
Therefore, as shown in Fig.~\ref{fig:fourdiagram2},
we conclude that the nonlocal resistance $R_{NL}$ consists of three terms:
\begin{eqnarray}
   R_{NL}=R_{ballistic}+R_{classic}+R_{Hall}.
\label{eq:ballistic}
\end{eqnarray}
Here, the first part $R_{ballistic}$ stands for the ballistic transport mechanism we discussed above.
The second part $R_{classic}$ represents the classical diffusion.
And the third part $R_{Hall}$ originates from the spin-orbit coupling term of Eq.~(\ref{eq:Hamiltonian}),
where the electron current flowing along the left vertical wires generates a perpendicular spin current
due to the SHE, and is finally converted to the electron current in the right vertical wires
due to the inverse spin Hall effect (ISHE).

Secondly, both the local and nonlocal resistance are symmetric about the line of $E_F=0$
and reach their maxima at $E_F=0$.
Importantly, similar to the findings in previous experiments, the nonlocal resistance $R_{NL}$
collapses to zero much more rapidly compared with the local resistance $R_L$.
In other words, the full width at half maximum (FWHM) of the nonlocal resistance
is much smaller than that of the local resistance.
For example, we can see from Fig.~\ref{fig:nonlocal1}(c) that the nonlocal resistance is decreased to nearly zero
and is maintained at about $R_{NL}=0.04h/e^2$ when $|E_F|>0.2$,
while the local resistance $R_L$ is still decreasing.
This phenomenon is contradictory to the known formula\cite{Abanin2}:
\begin{eqnarray}
   R_{NL}\propto\sigma_{xy}^2R_L^3,
\label{eq:Abanin}
\end{eqnarray}
where $\sigma_{xy}$ is the spin Hall conductance.
In fact, it seems that this contradiction can be partially explained by our analysis on the negative nonlocal resistance.
Specifically, the original value of the nonlocal resistance is $R_{Hall}$, which satisfies Eq.~(\ref{eq:Abanin}).
And $R_{classic}$ may result in a large peak near $E_F=0$ due to the extremely small DOS at the Dirac point.
Moreover, according to Eq.~(\ref{eq:ballistic}), there must exist an additional term $R_{ballistic}$ to $R_{NL}$.
Since the value of $R_{ballistic}$ is negative,
it is natural that $R_{NL}$ decays to zero much more rapidly than the theoretical prediction as shown Eq.~(\ref{eq:Abanin}).

Thirdly, according to the known literatures,
most of them attribute the fast decay of the nonlocal resistance to the SHE or the VHE.
Although our model is not exactly the same as those in the known literatures,
it is reasonable for us to make the same assumption.
However, by inspecting Fig.~\ref{fig:nonlocal1},
the only phenomenon we can find between the nonlocal resistance and the SHE is that the negative nonlocal resistance becomes weaker and weaker,
and gradually disappears with the increase of the spin-orbit coupling strength $\lambda_R$.
Correspondingly, the shrinking speed slows down as well.
Namely, we can only expect that the SHE mainly affects the value of $R_{NL}$ around $E_F=\pm0.1$.
Therefore, only Fig.~\ref{fig:nonlocal1} itself is not enough for us to fully understand the underlying mechanism
between the nonlocal resistance and the SHE.
Actually, we do not even know whether the fast decay of the nonlocal resistance
has any relationship to the SHE.
Therefore, it is necessary for us to obtain pure $R_{Hall}$ besides $R_{NL}$ shown in Fig.~\ref{fig:nonlocal1}.
One simple way is to take $R_{NL}$ in Fig.~\ref{fig:nonlocal1}(a) with $\lambda_R=0$ as a reference line, and we then
subtract this reference line from $R_{NL}(\lambda_R)$ to obtain $R_{Hall}(\lambda_R)=R_{NL}(\lambda_R)-R_{NL}(0)$.
However, after making this attempt, we find that the result is messy and disordered,
indicating that the ballistic transport and the classic diffusion are also affected by the SHE.
Therefore, we cannot simply take Fig.~\ref{fig:nonlocal1}(a) as a reference line to calculate $R_{Hall}$,
and we further consider a different system to investigate the Rashba effect in the next section.

\section{Newly designed six-terminal system to study the external Rashba effect}

\begin{figure}[h]
\includegraphics [width=\columnwidth]{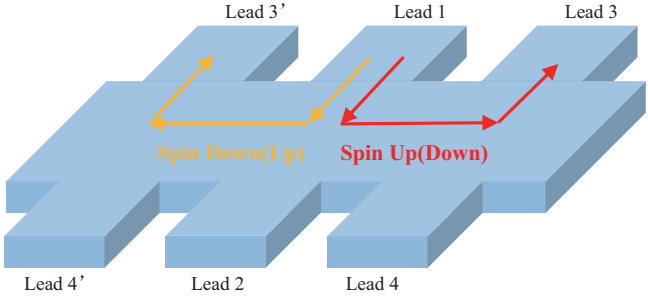}
\caption{(Color online)
The schematic diagram of the proposed six-terminal system.
Newly added lead 3' and lead 4' locate at a mirror symmetry to lead 3 and lead 4.
If the spin up (down) current injected into lead 1 transports along the red line,
the spin down (up) current must follow the yellow line.
}\label{fig:sixdiagram}
\end{figure}

In order to further study how the external Rashba effect affects the nonlocal resistance,
we investigate the transport properties of a six-terminal system, as shown in Fig.~\ref{fig:sixdiagram}, instead of the H-shaped four-terminal one.
The only difference between Fig.~\ref{fig:sixdiagram} and Fig.~\ref{fig:fourdiagram1} is that we add two leads (lead 3' and lead 4')
at the left side of lead 1 and lead 2, and keep them at the mirror sites of lead 3 and lead 4.
If a spin up current is injected into lead 1, e.g., by adding a ferromagnetic lead,
we can detect voltage signals $V_{34\uparrow}$ on lead 3 and lead 4.
Then the nonlocal resistance is written as:
\begin{eqnarray}
  R_{NL\uparrow}=R_{ballistic\uparrow}+R_{classic\uparrow}+R_{Hall\uparrow}.
\end{eqnarray}
Similarly, regarding a spin down current, the voltage $V_{34\downarrow}$ is detected
and the nonlocal resistance is expressed as:
\begin{eqnarray}
  R_{NL\downarrow}=R_{ballistic\downarrow}+R_{classic\downarrow}+R_{Hall\downarrow}.
\end{eqnarray}
Since the ballistic transport and the classic diffusion have no relationship to the spin direction,
the nonlocal resistance $R_{ballistic}$ and $R_{classic}$ caused by these two mechanisms
must be the same along the two different spin directions,
i.e., $R_{ballistic\uparrow}=R_{ballistic\downarrow}$ and $R_{classic\uparrow}=R_{classic\downarrow}$.
Therefore, with the detected $R_{NL\uparrow}$ and $R_{NL\downarrow}$,
we can easily remove the perturbations of the ballistic transport and the classic diffusion
based on this two-step proposal.
Finally, we can obtain the pure result caused by the Rashba effect as:
\begin{eqnarray}
   R_{Hall\uparrow}-R_{Hall\downarrow}=R_{NL\uparrow}-R_{NL\downarrow}.
\end{eqnarray}
Although we have not obtained $R_{Hall\uparrow\downarrow}$,
we will prove that $|R_{Hall\uparrow}-R_{Hall\downarrow}|$ itself
demonstrates the value of $R_{Hall\uparrow\downarrow}$ in the next paragraph.
In fact, the most advantage of this six-terminal system is that when the spin up current is injected into lead 1,
we can also detect voltage signals $V_{3'4'\uparrow}$ between leads 3' and 4', besides that between leads 3 and 4.
According to the symmetric analysis,
$V_{3'4'\uparrow}$ should be equal to the voltage $V_{34\downarrow}$ between leads 3 and 4 with the spin down current.
Therefore, we can easily obtain $R_{Hall\uparrow}-R_{Hall\downarrow}$ by one step without changing the magnetization direction of lead 1,
which is very important in the realistic experiments.
From now on, for clarity,
we always consider the spin currents along the two directions, though both the numerical calculations and experiments need only one kind of spin currents
in reality.

\begin{figure}[h]
\includegraphics [width=\columnwidth]{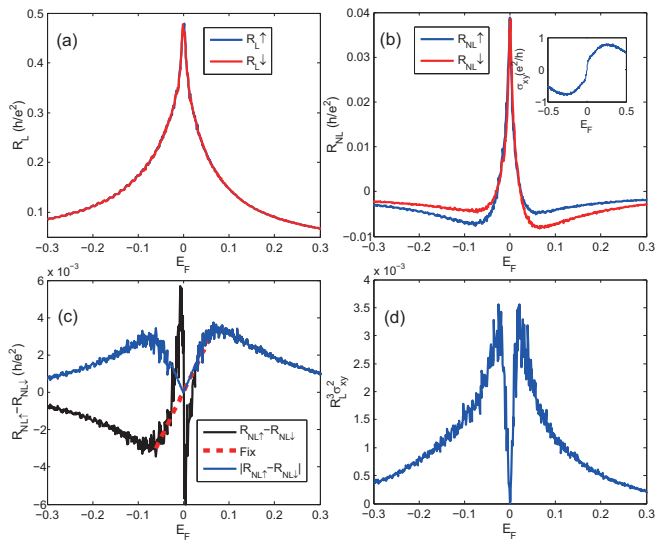}
\caption{(Color online)
(a) The local resistance $R_L$ for two spin directions.
(b) The nonlocal resistance $R_{NL}$ for two spin directions.
And the inset is the spin Hall conductance $\sigma_{xy}$.
(c) The black line is calculated based on $R_{NL\uparrow}-R_{NL\downarrow}$.
Since the numerical error results in dramatic oscillation of $R_{NL\uparrow}-R_{NL\downarrow}$ near $E_F=0$,
we add a red dashed line to describe the accurate behavior of $R_{NL\uparrow}-R_{NL\downarrow}$ near the Dirac point.
At last, the blue line represents $R_{Hall}$ that equals to the absolute value of the black line,
and the fix in red line is also considered.
(d) $R_{Hall}$ calculated based on Eq.~(\ref{eq:Abanin}).
}\label{fig:nonlocal2}
\end{figure}

In Fig.~\ref{fig:nonlocal2}(a), we first show the local resistance $R_L$ calculated according to the above proposal.
Since the local resistance has no relationship to the SHE and is not sensitive to the spin direction,
the local resistances in the two different spin directions exhibit almost the same behavior.
Then, in Fig.~\ref{fig:nonlocal2}(b), we show the nonlocal resistance $R_{NL\uparrow}$ and $R_{NL\downarrow}$,
which reflect the nonlocal resistance purely caused by the SHE.
It is clear that both $R_{NL\uparrow}$ colored by blue and $R_{NL\downarrow}$ colored by red are asymmetric.
$R_{NL\uparrow}$ with $E_F<0$ is smaller than that with $E_F>0$,
and $R_{NL\downarrow}$ shows the opposite behavior.
According to the discussion of the Hall effect part in Eq.~(\ref{eq:ballistic}),
the above phenomenon can be explained by the spin Hall conductance $\sigma_{xy}$\cite{Sheng},
which is calculated in a four-terminal system and is shown in the inset of Fig.~\ref{fig:nonlocal2}(b).
Specifically, $\sigma_{xy}$ is antisymmetric about the original point:
$\sigma_{xy}(E_F)=-\sigma_{xy}(-E_F)$ and $\sigma_{xy}>0$ when $E_F>0$.
As we all know, the sign reversing of $\sigma_{xy}$ denotes the direction conversing of the SHE.
That is to say,
if we assume that the spin up current turns left with a positive $\sigma_{xy}$ through the SHE
(other conditions can be analyzed similarly),
the spin up current injected into lead 1 will turn left as described by the red line in Fig.~\ref{fig:sixdiagram} when $E_F>0$.
Thus, the spin up current will contribute to $V_{34}$
and be reflected in $R_{NL}$ through the SHE.
This also tells us that, when $E_F<0$, the spin up current will turn right as descried by the yellow line in Fig.~\ref{fig:sixdiagram},
which means we can hardly detect its signal caused by the SHE on leads 3 and 4.
While for Fig.~\ref{fig:nonlocal2}(b), since $R_{NL\uparrow}$ of $E_F<0$ is smaller than that of $E_F>0$,
we can conclude that it is spin up current that turns left when $E_F>0$ as described by the red line in Fig.~\ref{fig:sixdiagram}.
Correspondingly, the spin down current will turn left when $E_F<0$, just as the red line in Fig.~\ref{fig:nonlocal2}(b).
Therefore, the nonlocal resistance for the spin up current in the region of $E_F<0$
and the spin down current for $E_F>0$ can be regarded as the reference line without the SHE.
In other words, we can obtain $R_{Hall}$ just by subtracting $R_{NL\uparrow}$ from $R_{NL\downarrow}$ when $E_F<0$ and subtracting $R_{NL\downarrow}$ from $R_{NL\uparrow}$ when $E_F>0$:
\begin{eqnarray}
   R_{Hall}=|R_{Hall\uparrow}-R_{Hall\downarrow}|=|R_{NL\uparrow}-R_{NL\downarrow}|.
\end{eqnarray}

Since Eq.~(\ref{eq:Abanin}) only considers the SHE without any other effects,
$R_{NL}$ in Eq.~(\ref{eq:Abanin}) should be replaced by $R_{Hall}$ actually.
In Fig.~\ref{fig:nonlocal2}(c), we first draw the nonlocal resistance $R_{Hall}$ colored in black based on $R_{Hall\uparrow}-R_{Hall\downarrow}$.
The drastic oscillation around $E_F=0$ mainly results from the numerical errors,
because the values of $R_{NL\uparrow\downarrow}$ around $E_F=0$ shown in Fig.~\ref{fig:nonlocal2}(b) are very sharp,
and a small error seems aggravating in Fig.~\ref{fig:nonlocal2}(c).
According to Fig.~\ref{fig:nonlocal2}(b), $\sigma_{xy}$ equals to zero at $E_F=0$, which means there exists no SHE at the Dirac point.
Thus, $R_{NL}$ should not be sensitive to the spin direction, and $R_{NL\uparrow}$ must be equal to $R_{NL\downarrow}$ at $E_F=0$.
Hence, the accurate value of $R_{Hall\uparrow}-R_{Hall\downarrow}$ should follow the red dashed line drawn in Fig.~\ref{fig:nonlocal2}(c),
which connects the two peaks and passes through the origin.
Then, in order to further justify our results,
we also calculate $R_{Hall}$
with the local resistance $R_L$ and the spin Hall conductance $\sigma_{xy}$ according to Eq.~(\ref{eq:Abanin}) in Fig.~\ref{fig:nonlocal2}(d),
and compare these two figures obtained by the two different methods.
As we can see, after folding Fig.~\ref{fig:nonlocal2}(c) by calculating the absolute value of $R_{Hall\uparrow}-R_{Hall\downarrow}$, as shown by the blue line in Fig.~\ref{fig:nonlocal2}(c),
the general behaviors of the two blue lines in Fig.~\ref{fig:nonlocal2}(c) and ~\ref{fig:nonlocal2}(d) look the same.
The minor difference may be caused by the proportional signal in Eq.~(\ref{eq:Abanin}),
which may contain parameters sensitive to the Fermi energy.
Importantly, we find that $R_{Hall}=0$ for $E_F=0$,
which means the peak of the nonlocal resistance shown in Fig.~\ref{fig:nonlocal1} has no relationship to the SHE.
In contrast to our previous prediction,
this phenomenon also means the rapid shrinking of $R_{NL}$ with the deviation of $E_F$ compared to $R_L$
is not caused by the Rashba effect.
Moreover, in Fig.~\ref{fig:nonlocal2}(c) and \ref{fig:nonlocal2}(d), we find there exist a pair of peaks around $E_F=\pm0.1$, which originates from the Rashba effect.
Thus, it is natural that the stronger the Rashba effect is, the more obvious the peaks become.
Now, we can explain the third feature of Fig.~\ref{fig:nonlocal1} in Sec.II.
As shown in Fig.~\ref{fig:nonlocal1}, the negative nonlocal resistance appears around $E_F=\pm0.1$,
which is exactly where the peaks of $R_{Hall}$ locate.
Thus, it is the peak of $R_{Hall}$ that counteracts the negative value of the nonlocal resistance $R_{NL}$.
As a result, we find that the negative value of $R_{NL}$ gradually disappears
with the increasing Rashba effect, as shown in Fig.~\ref{fig:nonlocal1}.

To summarize, with the calculated $R_{Hall}$ and further analysis,
we first conclude that the large peak of the nonlocal resistance at $E_F=0$ has no relationship to the Rashba effect.
Actually, it is possible that this large peak mainly originates from the extremely small DOS of the monolayer graphene at the Dirac point.
Similar to our results, not long ago, several groups also doubted the direct connection between the peak of $R_{NL}$ and the SHE as assumed in early papers\cite{Wang2015,Kaverzin2015,Tuan2016}.
Then, we know that the fast decay of the nonlocal resistance is not caused by the Rashaba effect as predicted.
Considering Sec.II, it is the negative $R_{ballistic}$ caused by the ballistic transport
that leads to this interesting phenomenon.
Moreover, since there exists one pair of $R_{Hall}$ peaks at $E_F=\pm0.1$
where the negative dip of $R_{ballistic}$ locates,
the Rashba effect itself actually plays a negative role to the fast decay of the nonlocal resistance.
That is also why we find the tendency of the rapid shrinking of $R_{NL}$ becomes weaker and weaker in Fig.~\ref{fig:nonlocal1}.

\section{Conclusion and discussion}
In conclusion, using the non-equilibrium Green's function method,
we obtain the local and nonlocal resistance in an H-shaped graphene, similar to the real experiments.
Specifically, there does exist a large peak of the nonlocal resistance $R_{NL}$ at the Dirac point.
In particular, we do find $R_{NL}$ decreasing much more quickly than $R_L$
when the Fermi energy deviates from the Dirac point.
Besides, we have proven that the total nonlocal resistance $R_{NL}$ stems from three kinds of mechanisms:
the ballistic transport $R_{ballistic}$, the classic diffusion $R_{classic}$ and the SHE $R_{Hall}$.
After a further calculation of a six-terminal system,
we conclude that the peak of $R_{NL}$ and its rapid decrease do not result from the Rashba effect,
but originate from the small DOS near the Dirac point and the ballistic transport, respectively.
Moreover, the Rashba effect itself actually plays a negative role in this rapid decrease.
The whole physical pictures behind can be concluded as:
first, because of the extremely small DOS, there exists a giant peak of the nonlocal resistance at the Dirac point;
then, due to the ballistic transport mechanism, the negative value of $R_{ballistic}$ leads to the fast decay of $R_{NL}$;
and finally, $R_{Hall}$ originating from the SHE will offset this fast decay somewhat.

Though lots of experiments found the signal of giant nonlocal resistance $R_{NL}$,
which decreases much more rapidly compared with the local resistance $R_{L}$, in an H-shaped graphene,
and attributed them to the spin/valley Hall effect,
we give a numerical simulation presenting an explanation different from the previous prediction.
Here, we have to emphasize that this difference does not mean that all previous conclusions obtained from the experiments are incorrect,
because the model, the sample size and the spin-orbit coupling we use might be different.
Finally, the six-terminal method proposed in this work
is very helpful to study the underlying mechanism of transport with the SHE,
because the SHE itself is always mixed with
and not easy to be separated from other mechanisms, such as the classical diffusion etc.

\section*{ACKNOWLEDGMENTS}
We thank the insightful discussions with Jie Liu, Ai-min Guo, Qing-feng Sun, Jianhao Chen, Xi Lin and Wei Han.
This work was financially supported by NBRPC (Grant No. 2015CB921102, 2014CB920901)
and NSFC (Grants Nos. 11534001, 11374219, 11504008).


\begin{thebibliography}{10}
\bibitem{Andreas}
\bibinfo{author}{Andreas Roth, Christoph Br\"{u}ne, Hartmut Buhmann, Laurens W. Molenkamp, Joseph Maciejko,
Xiao-Liang Qi, and Shou-Cheng Zhang},
\bibinfo{journal}{Science}
\textbf{\bibinfo{volume}{325}}, \bibinfo{pages}{294}
(\bibinfo{year}{2009}).

\bibitem{Chang}
\bibinfo{author}{Cui-Zu Chang, Weiwei Zhao, Duk Y. Kim, Peng Wei, J.K. Jain, Chaoxing Liu, Moses H.W. Chan, and Jagadeesh S. Moodera},
\bibinfo{journal}{Phys. Rev. Lett.}
\textbf{\bibinfo{volume}{115}}, \bibinfo{pages}{057206}
(\bibinfo{year}{2015}).

\bibitem{Parameswaran}
\bibinfo{author}{S. A. Parameswaran, T. Grover, D. A. Abanin, D. A. Pesin, and A. Vishwanath},
\bibinfo{journal}{Phys. Rev. X}
\textbf{\bibinfo{volume}{4}}, \bibinfo{pages}{031035}
(\bibinfo{year}{2014}).

\bibitem{McEuen}
\bibinfo{author}{P. L. McEuen, A. Szafer, C. A. Richter, B.W. Alphenaar,
J. K. Jain, A. D. Stone, R. G. Wheeler, and R. N. Sacks},
\bibinfo{journal}{Phys. Rev. Lett.}
\textbf{\bibinfo{volume}{64}}, \bibinfo{pages}{2062}
(\bibinfo{year}{1990}).

\bibitem{Abanin1}
\bibinfo{author}{D. A. Abanin, S. V. Morozov, L. A. Ponomarenko, R. V. Gorbachev, A. S. Mayorov,
M. I. Katsnelson, K. Watanabe, T. Taniguchi, K. S. Novoselov, L. S. Levitov, and A. K. Geim},
\bibinfo{journal}{Science}
\textbf{\bibinfo{volume}{332}}, \bibinfo{pages}{328}
(\bibinfo{year}{2011}).

\bibitem{Balakrishnan}
\bibinfo{author}{Jayakumar Balakrishnan, Gavin KokWai Koon, Manu Jaiswal, A. H. Castro Neto, and Barbaros \"{O}zyilmaz},
\bibinfo{journal}{Nat. Phys.}
\textbf{\bibinfo{volume}{9}}, \bibinfo{pages}{284}
(\bibinfo{year}{2013}).


\bibitem{Gorbachev}
\bibinfo{author}{R. V. Gorbachev, J. C. W. Song, G. L. Yu, A. V. Kretinin, F. Withers, Y. Cao,
A. Mishchenko, I. V. Grigorieva, K. S. Novoselov, L. S. Levitov, and A. K. Geim},
\bibinfo{journal}{Science}
\textbf{\bibinfo{volume}{346}}, \bibinfo{pages}{448}
(\bibinfo{year}{2014}).

\bibitem{Shimazaki}
\bibinfo{author}{Y. Shimazaki, M. Yamamoto,	I. V. Borzenets, K. Watanabe, T. Taniguchi, and S. Tarucha},
\bibinfo{journal}{Nat. Phys.}
\textbf{\bibinfo{volume}{11}}, \bibinfo{pages}{1032-1036}
(\bibinfo{year}{2015}).

\bibitem{Sui}
\bibinfo{author}{Mengqiao Sui, Guorui Chen,	Liguo Ma, Wen-Yu Shan, Dai Tian, Kenji Watanabe, Takashi Taniguchi,	Xiaofeng Jin, Wang Yao,	Di Xiao, and Yuanbo Zhang},
\bibinfo{journal}{Nat. Phys.}
\textbf{\bibinfo{volume}{11}}, \bibinfo{pages}{1027-1031}
(\bibinfo{year}{2015}).

\bibitem{Michihisa}
\bibinfo{author}{Michihisa Yamamoto, Yuya Shimazaki, Ivan V. Borzenets, and Seigo Tarucha},
\bibinfo{journal}{J. Phys. Soc. Jpn}
\textbf{\bibinfo{volume}{84}}, \bibinfo{pages}{121006}
(\bibinfo{year}{2015}).

\bibitem{Hirsch}
\bibinfo{author}{J. E. Hirsch},
\bibinfo{journal}{Phys. Rev. Lett.}
\textbf{\bibinfo{volume}{83}}, \bibinfo{pages}{1834}
(\bibinfo{year}{1999}).

\bibitem{Murakami}
\bibinfo{author}{S. Murakami, N. Nagaosa, and S. C. Zhang},
\bibinfo{journal}{Science}
\textbf{\bibinfo{volume}{301}}, \bibinfo{pages}{1348}
(\bibinfo{year}{2003}).

\bibitem{Sinova}
\bibinfo{author}{J. Sinova, D. Culcer, Q. Niu, N. A. Sinitsyn, T. Jungwirth, and A. H. MacDonald},
\bibinfo{journal}{Phys. Rev. Lett.}
\textbf{\bibinfo{volume}{92}}, \bibinfo{pages}{126603}
(\bibinfo{year}{2004}).

\bibitem{Kato}
\bibinfo{author}{Y. K. Kato, R. C. Myers, A. C. Gossard, and D. D. Awsschalom},
\bibinfo{journal}{Science}
\textbf{\bibinfo{volume}{306}}, \bibinfo{pages}{1910}
(\bibinfo{year}{2004}).

\bibitem{Kimura}
\bibinfo{author}{T. Kimura and Y. Otani},
\bibinfo{journal}{Phys. Rev. Lett.}
\textbf{\bibinfo{volume}{99}}, \bibinfo{pages}{196604}
(\bibinfo{year}{2007}).

\bibitem{Brune}
\bibinfo{author}{C. Br\"{u}ne, A. Roth, E. G. Novik, M. K\"{o}nig, H. Buhmann, E. M. Hankiewicz, W. Hanke, J. Sinova, and L. W. Molenkamp},
\bibinfo{journal}{Nat. Phys.}
\textbf{\bibinfo{volume}{6}}, \bibinfo{pages}{448-454}
(\bibinfo{year}{2010}).




\bibitem{Abanin2}
\bibinfo{author}{D. A. Abanin, A. V. Shytov, L. S. Levitov, and B. I. Halperin},
\bibinfo{journal}{Phys. Rev. B}
\textbf{\bibinfo{volume}{79}}, \bibinfo{pages}{035304}
(\bibinfo{year}{2009}).

\bibitem{Chen}
J.H. Chen, private communication, and the experimental work will be online soon.

\bibitem{Mihajlovic}
\bibinfo{author}{G. Mihajlovic, J. E. Pearson, M. A. Garcia, S. D. Bader, and A. Hoffmann},
\bibinfo{journal}{Phys. Rev. Lett.}
\textbf{\bibinfo{volume}{103}}, \bibinfo{pages}{166601}
(\bibinfo{year}{2009}).

\bibitem{foot1}
The calculation process is actually based on a six-terminal system.
That is to say, there exist additional two leads at the left and right side of the center region, marked by lead L and lead R,
because the definition of $R_L$ in some experiments requires these two leads.
However, the deduction of the nonlocal resistance $R_{NL}$ and the local resistance $R_L$ in our paper
doesn't need any information from lead L and lead R.
Therefore, we just claim that it is a four-terminal system that we use for simplicity.

\bibitem{book}Electronic Transport in Mesoscopic Systems, edited by S. Datta (Cambridge University Press, Cambridge, England, 1995).

\bibitem{Jiang}
\bibinfo{author}{Hua Jiang, Lei Wang, Qing-feng Sun, and X. C. Xie},
\bibinfo{journal}{Phys. Rev. B}
\textbf{\bibinfo{volume}{80}}, \bibinfo{pages}{165316}
(\bibinfo{year}{2009}).

\bibitem{footnote}
Actually, we have also calculated another condition with $M=100$,
which seems more like the real experiments.
However, the results are nearly the same between $M=30$ and $M=100$, except the order of magnitude.
Therefore, we just show the condition with $M=30$ in order to simplify our calculation.


\bibitem{Sheng}
\bibinfo{author}{L. Sheng, D. N. Sheng, and C. S. Ting},
\bibinfo{journal}{Phys. Rev. Lett.}
\textbf{\bibinfo{volume}{94}}, \bibinfo{pages}{016602}
(\bibinfo{year}{2005}).

\bibitem{Wang2015}
\bibinfo{author}{Y. Wang, X. Cai, J. Reutt-Robey, and M. S. Fuhrer},
\bibinfo{journal}{Phys. Rev. B}
\textbf{\bibinfo{volume}{92}}, \bibinfo{pages}{161411}
(\bibinfo{year}{2015}).

\bibitem{Kaverzin2015}
\bibinfo{author}{A. A. Kaverzin and B. J. van Wees},
\bibinfo{journal}{Phys. Rev. B}
\textbf{\bibinfo{volume}{91}}, \bibinfo{pages}{165412}
(\bibinfo{year}{2015}).

\bibitem{Tuan2016}
\bibinfo{author}{Dinh Van Tuan, J. M. Marmolejo-Tejada, Xavier Waintal, Branislav K. Nikoli\'{c}, and Stephan Roche},
\bibinfo{journal}{arXiv:1603.03870v1}
\end{thebibliography}
\end{document}